\documentclass[italian,english]{article}
\usepackage[T1]{fontenc}
\usepackage[latin1]{inputenc}
\usepackage{float}
\usepackage{color}
\usepackage{graphicx}
\usepackage{amssymb}

\makeatletter

 \newcommand{\lyxaddress}[1]{
   \par {\raggedright #1 
   \vspace{1.4em}
   \noindent\par}
 }

\usepackage{babel}
\makeatother
\begin{document}

\title{\textbf{Primordial production of massive relic gravitational waves
from a weak modification of General Relativity}}

\author{\textbf{Christian Corda }}

\maketitle

\lyxaddress{\begin{center}Centro di Scienze Naturali, Via di Galceti 74 - 59100
PRATO, Italy and Associazione Galileo Galilei, Via Pier Cironi 16
- 59100 PRATO, Italy \end{center}}

\begin{center}\textit{E-mail address:} \textcolor{blue}{christian.corda@ego-gw.it} \end{center}

\begin{abstract}
The production of a stochastic background of relic gravitational waves
is well known in various works in the literature, where, using the
so called adiabatically-amplified zero-point fluctuations process
it has been shown how the standard inflationary scenario for the early
universe can in principle provide a distinctive spectrum of relic
gravitational waves. In this paper, it is shown that a weak modification
of General Relativity produces a third massive polarization of gravitational
waves and the primordial production of this polarization is analysed
adapting the adiabatically-amplified zero-point fluctuations process
at this case. 

The presence of the mass could also have important applications in
cosmology as the fact that gravitational waves can have mass could
give a contribution to the dark matter of the Universe. 

At the end of the paper an upper bound for these relic gravitational
waves, which arises from the WMAP constrains, is also released.
\end{abstract}

\section{Introduction}

Recently, the data analysis of interferometric gravitational waves
(GWs) detectors has been started (for the current status of GWs interferometers
see \cite{key-1,key-2,key-3,key-4,key-5,key-6,key-7,key-8}) and the
scientific community aims in a first direct detection of GWs in next
years. 

Detectors for GWs will be important for a better knowledge of the
Universe and also to confirm or ruling out the physical consistency
of General Relativity or of any other theory of gravitation \cite{key-9,key-10,key-11,key-12,key-13,key-14}.
This is because, in the context of Extended Theories of Gravity, some
differences between General Relativity and the others theories can
be pointed out starting by the linearized theory of gravity \cite{key-9,key-10,key-12,key-14}.
In this picture, detectors for GWs are in principle sensitive also
to a hypotetical \textit{scalar} component of gravitational radiation,
that appears in extended theories of gravity like scalar-tensor gravity
and high order theories \cite{key-12,key-15,key-16,key-17,key-18,key-19,key-20,key-21,key-22},
Brans-Dicke theory \cite{key-23} and string theory \cite{key-24}.

A possible target of these experiments is the so called stochastic
background of gravitational waves \cite{key-25,key-26,key-27,key-28,key-29,key-30}. 

The production of the primordial part of this stochastick background
(relic GWs) is well known in the literature starting by the works
of \cite{key-25,key-26} and \cite{key-27,key-28}, that, using the
so called adiabatically-amplified zero-point fluctuations process,
have shown in two different ways how the standard inflationary scenario
for the early universe can in principle provide a distinctive spectrum
of relic gravitational waves, while in \cite{key-29,key-30} the primordial
production has been analyzed for the scalar component admitted from
scalar-tensor gravity. In this paper, it is shown that a weak modification
of General Relativity generates a third massive polarization of gravitational
waves and the primordial production of this polarization is analysed
adapting the adiabatically-amplified zero-point fluctuations process
at this case. We have also to emphasize that, in a recent paper \cite{key-40},
such a process has been applied to the same theory which we are going
to discuss in the present work. But, in \cite{key-40} a different
point of view has been considered. In that case, using a conform analysis,
the authors discussed such a process in respect to the two standard
polarizations which arises from standard General Relativity. In the
present paper the analysis is focused to the third massive polarization.

The presence of the mass could also have important applications in
cosmology because the fact that gravitational waves can have mass
could give a contribution to the dark matter of the Universe. 

At the end of the paper an upper bound for these relic gravitational
waves, which arises from the WMAP constrains, is also released.

\section{A weak modification of general relativity}

Let us consider the action\begin{equation}
S=\int d^{4}x\sqrt{-g}f_{0}R^{1+\varepsilon}+\mathcal{L}_{m}\label{eq: high order 1}\end{equation}

Equation (\ref{eq: high order 1}) is a particular choice in $f(R)$
theories of gravity \cite{key-9,key-10,key-11,key-13,key-19,key-21}
in respect to the well known canonical one of General Relativity (the
Einstein - Hilbert action \cite{key-31,key-32}) which is 

\begin{equation}
S=\int d^{4}x\sqrt{-g}R+\mathcal{L}_{m}.\label{eq: EH}\end{equation}

Criticisms on $f(R)$ theories of gravity arises from the fact that
lots of such theories can be excluded by requirements of Cosmology
and Solar System tests \cite{key-33}. But, in the case of the action
(\ref{eq: high order 1}), the variation from standard General Relativity
is very weak, becauese $\varepsilon$ is a very small real parameter,
thus, the mentioned constrains could be, in principle, satisfed \cite{key-33}.
Note: General Relativity is obtained for $\varepsilon=0$ and $f_{0}=1.$

The action (\ref{eq: high order 1}) has been analyzed in \cite{key-34}
in a cosmologic context. But, because we will interact with gravitational
waves, i.e. the linearized theory in vacuum, $\mathcal{L}_{m}=0$
will be put and the pure curvature action \begin{equation}
S=\int d^{4}x\sqrt{-g}f_{0}R^{1+\varepsilon}\label{eq: high order 12}\end{equation}
 will be considered.

\section{The field equations}

Following \cite{key-32,key-35} (note that in this paper we work with
$8\pi G=1$, $c=1$ and $\hbar=1$), the variational principle

\begin{equation}
\delta\int d^{4}x\sqrt{-g}f_{0}R^{1+\varepsilon}=0\label{eq: high order 2}\end{equation}

in a local Lorentz frame will be used.

One gets: \begin{equation}
\begin{array}{c}
\delta\int d^{4}x\sqrt{-g}f_{0}R^{1+\varepsilon}=\int d^{4}x[\delta\sqrt{-g}f_{0}R^{1+\varepsilon}+f_{0}\sqrt{-g}\delta R^{1+\varepsilon}]=\\
\\=\int d^{4}x[\sqrt{-g}f_{0}(1+\varepsilon)R^{\varepsilon}R_{\mu\nu}-\frac{1}{2}g_{\mu\nu}f_{0}R^{1+\varepsilon}]\delta g^{\mu\nu}+\\
\\+\int d^{4}x\sqrt{-g}(1+\varepsilon)f_{0}R^{\varepsilon}g^{\mu\nu}\delta R_{\mu\nu}.\end{array}\label{eq: high order 3}\end{equation}

Recalling the relation between the Christoffel coefficients and the
Ricci tensor \cite{key-32,key-35} one can write

\begin{equation}
g^{\mu\nu}\delta R_{\mu\nu}=g^{\mu\nu}\partial_{\alpha}(\delta\Gamma_{\mu\nu}^{\alpha})-g^{\mu\alpha}\partial_{\alpha}(\delta\Gamma_{\mu\nu}^{\nu})\equiv\partial_{\alpha}X^{\alpha},\label{eq: X}\end{equation}

where \begin{equation}
X^{\alpha}\equiv g^{\mu\nu}(\delta\Gamma_{\mu\nu}^{\alpha})-g^{\mu\alpha}(\delta\Gamma_{\mu\nu}^{\nu}).\label{eq: di X}\end{equation}

In this way, the second integral in equation (\ref{eq: high order 3})
can be computed as

\begin{equation}
\begin{array}{c}
\int d^{4}x\sqrt{-g}(1+\varepsilon)f_{0}R^{\varepsilon}g^{\mu\nu}\delta R_{\mu\nu}=\int d^{4}x\sqrt{-g}(1+\varepsilon)f_{0}R^{\varepsilon}\partial_{\alpha}X^{\alpha}=\\
\\=\int d^{4}x\partial_{\alpha}[\sqrt{-g}(1+\varepsilon)f_{0}R^{\varepsilon}X^{\alpha}]-\int d^{4}x\partial_{\alpha}[\sqrt{-g}(1+\varepsilon)f_{0}R^{\varepsilon}]X^{\alpha}.\end{array}\label{eq: calcolo}\end{equation}

Assuming that fields are equal to zero at infinity \cite{key-32,key-35},
one gets 

\begin{equation}
d^{4}x\sqrt{-g}(1+\varepsilon)f_{0}R^{\varepsilon}g^{\mu\nu}\delta R_{\mu\nu}=-\int d^{4}x\partial_{\alpha}[\sqrt{-g}((1+\varepsilon)f_{0}R^{\varepsilon}]X^{\alpha}.\label{calcolo 2}\end{equation}

Now, let us compute $X^{\alpha}.$ Recalling that in a local Lorentz
frame it is

\begin{equation}
\bigtriangledown_{\beta}g_{\mu\nu}=\partial_{\beta}g_{\mu\nu}=0\label{eq: delta-delta}\end{equation}

and using the well known definitions of the Christofell coefficients
\cite{key-32,key-35} it is

\begin{equation}
\begin{array}{c}
\delta\Gamma_{\mu\nu}^{\alpha}=\delta[\frac{1}{2}g^{\beta\alpha}(\partial_{\mu}g_{\beta\nu}+\partial_{\nu}g_{\mu\beta}-\partial_{\beta}g_{\mu\nu})]=\\
\\=\frac{1}{2}g^{\beta\alpha}(\partial_{\mu}\delta g_{\beta\nu}+\partial_{\nu}\delta g_{\mu\beta}-\partial_{\beta}\delta g_{\mu\nu}).\end{array}\label{eq: g}\end{equation}

In the same way it is \begin{equation}
\delta\Gamma_{\mu\nu}^{\nu}=\frac{1}{2}g^{\nu\beta}\partial_{\mu}(\delta g_{\nu\beta}).\label{eq: g2}\end{equation}

From eqs. (\ref{eq: g}) and (\ref{eq: g2}) one gets

\begin{equation}
g^{\mu\nu}(\delta\Gamma_{\mu\nu}^{\alpha})=\frac{1}{2}\partial^{\alpha}(g_{\mu\nu}\delta g^{\mu\nu})-\partial^{\mu}(g_{\beta\mu}\delta g^{\nu\beta})\label{eq: calcola}\end{equation}

and \begin{equation}
g^{\mu\alpha}(\delta\Gamma_{\mu\nu}^{\nu})=-\frac{1}{2}\partial^{\alpha}(g_{\nu\beta}\delta g^{\nu\beta}).\label{eq: calcola 2}\end{equation}

Then, substituting in (\ref{eq: di X}), it is\begin{equation}
X^{\alpha}=\partial^{\alpha}(g_{\mu\nu}\delta g^{\mu\nu})-\partial^{\mu}(g_{\mu\nu}\delta g^{\alpha\nu}).\label{eq: X2}\end{equation}
With this equation, equation (\ref{calcolo 2}) becomes \begin{equation}
\begin{array}{c}
\int d^{4}x\sqrt{-g}(1+\varepsilon)f_{0}R^{\varepsilon}g^{\mu\nu}\delta R_{\mu\nu}=\\
\\=\int d^{4}x\partial_{\alpha}[\sqrt{-g}(1+\varepsilon)f_{0}R^{\varepsilon}][\partial^{\mu}(g_{\mu\nu}\delta g^{\alpha\nu})-\partial^{\alpha}(g_{\mu\nu}\delta g^{\mu\nu})],\end{array}\label{eq: calcolo 3}\end{equation}

which also gives \begin{equation}
\begin{array}{c}
\int d^{4}x\sqrt{-g}(1+\varepsilon)f_{0}R^{\varepsilon}g^{\mu\nu}\delta R_{\mu\nu}=\\
\\=\int d^{4}x\{ g_{\mu\nu}\partial^{\alpha}\partial_{\alpha}[\sqrt{-g}(1+\varepsilon)f_{0}R^{\varepsilon}]\delta g^{\mu\nu}\}-\int d^{4}x\{ g_{\mu\nu}\partial^{\mu}\partial_{\alpha}[\sqrt{-g}(1+\varepsilon)f_{0}R^{\varepsilon}]\delta g^{\alpha\nu}\}.\end{array}\label{eq: calcolo 4}\end{equation}

Putting this equation in the variation (\ref{eq: high order 3}) one
obtains \begin{equation}
\begin{array}{c}
\delta\int d^{4}x\sqrt{-g}(1+\varepsilon)f_{0}R^{\varepsilon}=\int d^{4}x[\sqrt{-g}(1+\varepsilon)f_{0}R^{\varepsilon}R_{\mu\nu}-\frac{1}{2}g_{\mu\nu}f_{0}R^{1+\varepsilon}]\delta g^{\mu\nu}+\\
\\+\int d^{4}x\{ g_{\mu\nu}\partial^{\alpha}\partial_{\alpha}[\sqrt{-g}(1+\varepsilon)f_{0}R^{\varepsilon})]-g_{\alpha\nu}\partial^{\mu}\partial_{\alpha}[\sqrt{-g}(1+\varepsilon)f_{0}R^{\varepsilon}]\}\delta g^{\mu\nu}\}.\end{array}\label{eq: high order fin}\end{equation}

The above variation is equal to zero for

\begin{equation}
(1+\varepsilon)f_{0}R^{\varepsilon}R_{\mu\nu}-\frac{1}{2}g_{\mu\nu}f_{0}R^{1+\varepsilon}=(\bigtriangledown_{\mu}\bigtriangledown_{\nu}-g_{\mu\nu}\square)(1+\varepsilon)f_{0}R^{\varepsilon},\label{eq: einstein}\end{equation}

which are the modified Einstein field equations. Writing down, exlplicitly,
the Einstein tensor eqs. (\ref{eq: einstein}) become\begin{equation}
G_{\mu\nu}=\frac{1}{(1+\varepsilon)f_{0}R^{\varepsilon}}\{-\frac{1}{2}g_{\mu\nu}\varepsilon f_{0}R^{1+\varepsilon}+[(1+\varepsilon)f_{0}R^{\varepsilon}]_{;\mu;\nu}-g_{\mu\nu}\square[(1+\varepsilon)f_{0}R^{\varepsilon}]\}.\label{eq: einstein 2}\end{equation}

Taking the trace of the field equations (\ref{eq: einstein 2}) one
gets 

\begin{equation}
\square(1+\varepsilon)f_{0}R^{\varepsilon}=\frac{(1-\varepsilon)}{3}f_{0}R^{1+\varepsilon}.\label{eq: KG}\end{equation}

Now, we can define the \textit{effective} scalar field \begin{equation}
\Phi\equiv(1+\varepsilon)f_{0}R^{\varepsilon}\label{eq: campo effettivo}\end{equation}

with associated an effective potential

\begin{equation}
\frac{dV}{d\Phi}\equiv\frac{(1-\varepsilon)}{3}f_{0}R^{1+\varepsilon}.\label{eq: potenziale effettivo}\end{equation}

Thus, from eq. (\ref{eq: KG}), a Klein - Gordon equation for the
effective $\Phi$ scalar field is obtained:

\begin{equation}
\square\Phi=\frac{dV}{d\Phi}.\label{eq: KG2}\end{equation}

\section{The linearized theory}

To study gravitational waves, the linearized theory has to be analyzed,
with a little perturbation of the background, which is assumed given
by a near Minkowskian background, i.e. a Minkowskian background plus
$\Phi=\Phi_{0}$ (the Ricci scalar is assumed constant in the background)
\cite{key-9,key-19}. We also assume $\Phi_{0}$ to be a minimum for
the effective potential $V$: 

\begin{equation}
V\simeq\frac{1}{2}\alpha\delta\Phi^{2}\Rightarrow\frac{dV}{d\Phi}\simeq m^{2}\delta\Phi,\label{eq: minimo}\end{equation}

and the constant $m$ has mass dimension. 

Putting

\begin{equation}
\begin{array}{c}
g_{\mu\nu}=\eta_{\mu\nu}+h_{\mu\nu}\\
\\\Phi=\Phi_{0}+\delta\Phi.\end{array}\label{eq: linearizza}\end{equation}

to first order in $h_{\mu\nu}$ and $\delta\Phi$, calling $\widetilde{R}_{\mu\nu\rho\sigma}$
, $\widetilde{R}_{\mu\nu}$ and $\widetilde{R}$ the linearized quantity
which correspond to $R_{\mu\nu\rho\sigma}$ , $R_{\mu\nu}$ and $R$,
the linearized field equations are obtained \cite{key-12,key-19,key-31}:

\begin{equation}
\begin{array}{c}
\widetilde{R}_{\mu\nu}-\frac{\widetilde{R}}{2}\eta_{\mu\nu}=(\partial_{\mu}\partial_{\nu}h_{m}-\eta_{\mu\nu}\square h_{m})\\
\\{}\square h_{m}=m^{2}h_{m},\end{array}\label{eq: linearizzate1}\end{equation}

where 

\begin{equation}
h_{m}\equiv\frac{\delta\Phi}{\Phi_{0}}.\label{eq: definizione}\end{equation}

Then, from the second of eqs. (\ref{eq: linearizzate1}), one can
define the mass like

\begin{equation}
m\equiv\sqrt{\frac{\square h_{m}}{h_{m}}}=\sqrt{\frac{\square\delta\Phi}{\delta\Phi}}=\sqrt{\frac{\square\delta R^{\varepsilon}}{\delta R^{\varepsilon}}}.\label{eq: massa}\end{equation}

Thus, as the mass is generated by variation of the Ricci scalar, we
can say that, in a certain sense, it is generated by variation of
spacetime curvature, re-obtaining the same result of \cite{key-9,key-19}.
The difference with the works \cite{key-9,key-19} is that now the
theory is more suitable as the modification of General Relativity
is very weak and in agreement with requirements of Cosmology and Solar
System tests \cite{key-33}. 

$\widetilde{R}_{\mu\nu\rho\sigma}$ and eqs. (\ref{eq: linearizzate1})
are invariants for gauge transformations \cite{key-9,key-12,key-19}

\begin{equation}
\begin{array}{c}
h_{\mu\nu}\rightarrow h'_{\mu\nu}=h_{\mu\nu}-\partial_{(\mu}\epsilon_{\nu)}\\
\\\delta\Phi\rightarrow\delta\Phi'=\delta\Phi;\end{array}\label{eq: gauge}\end{equation}

then 

\begin{equation}
\bar{h}_{\mu\nu}\equiv h_{\mu\nu}-\frac{h}{2}\eta_{\mu\nu}+\eta_{\mu\nu}h_{m}\label{eq: ridefiniz}\end{equation}

can be defined, and, considering the transform for the parameter $\epsilon^{\mu}$

\begin{equation}
\square\epsilon_{\nu}=\partial^{\mu}\bar{h}_{\mu\nu},\label{eq:lorentziana}\end{equation}
 a gauge parallel to the Lorenz one of electromagnetic waves can be
choosen:

\begin{equation}
\partial^{\mu}\bar{h}_{\mu\nu}=0.\label{eq: cond lorentz}\end{equation}

In this way field equations read like

\begin{equation}
\square\bar{h}_{\mu\nu}=0\label{eq: onda T}\end{equation}

\begin{equation}
\square h_{m}=m^{2}h_{m}\label{eq: onda S}\end{equation}

Solutions of eqs. (\ref{eq: onda T}) and (\ref{eq: onda S}) are
plan waves \cite{key-12,key-19}:

\begin{equation}
\bar{h}_{\mu\nu}=A_{\mu\nu}(\overrightarrow{p})\exp(ip^{\alpha}x_{\alpha})+c.c.\label{eq: sol T}\end{equation}

\begin{equation}
h_{m}=a(\overrightarrow{p})\exp(iq^{\alpha}x_{\alpha})+c.c.\label{eq: sol S}\end{equation}

where

\begin{equation}
\begin{array}{ccc}
k^{\alpha}\equiv(\omega,\overrightarrow{p}) &  & \omega=p\equiv|\overrightarrow{p}|\\
\\q^{\alpha}\equiv(\omega_{m},\overrightarrow{p}) &  & \omega_{m}=\sqrt{m^{2}+p^{2}}.\end{array}\label{eq: k e q}\end{equation}

In eqs. (\ref{eq: onda T}) and (\ref{eq: sol T}) the equation and
the solution for the standard waves of General Relativity \cite{key-31,key-32}
have been obtained, while eqs. (\ref{eq: onda S}) and (\ref{eq: sol S})
are respectively the equation and the solution for the massive mode
(see also \cite{key-9,key-12,key-19}).

The fact that the dispersion law for the modes of the massive field
$h_{m}$ is not linear has to be emphatized. The velocity of every
{}``ordinary'' (i.e. which arises from General Relativity) mode
$\bar{h}_{\mu\nu}$ is the light speed $c$, but the dispersion law
(the second of eq. (\ref{eq: k e q})) for the modes of $h_{m}$ is
that of a massive field which can be discussed like a wave-packet
\cite{key-9,key-12,key-19}. Also, the group-velocity of a wave-packet
of $h_{m}$ centered in $\overrightarrow{p}$ is 

\begin{equation}
\overrightarrow{v_{G}}=\frac{\overrightarrow{p}}{\omega},\label{eq: velocita' di gruppo}\end{equation}

which is exactly the velocity of a massive particle with mass $m$
and momentum $\overrightarrow{p}$.

From the second of eqs. (\ref{eq: k e q}) and eq. (\ref{eq: velocita' di gruppo})
it is simple to obtain:

\begin{equation}
v_{G}=\frac{\sqrt{\omega^{2}-m^{2}}}{\omega}.\label{eq: velocita' di gruppo 2}\end{equation}

Then, wanting a constant speed of the wave-packet, it has to be \cite{key-9,key-12,key-19}

\begin{equation}
m=\sqrt{(1-v_{G}^{2})}\omega.\label{eq: relazione massa-frequenza}\end{equation}

The relation (\ref{eq: relazione massa-frequenza}) is shown in fig.
1 for a value $v_{G}=0.9$.

\begin{figure}
\includegraphics{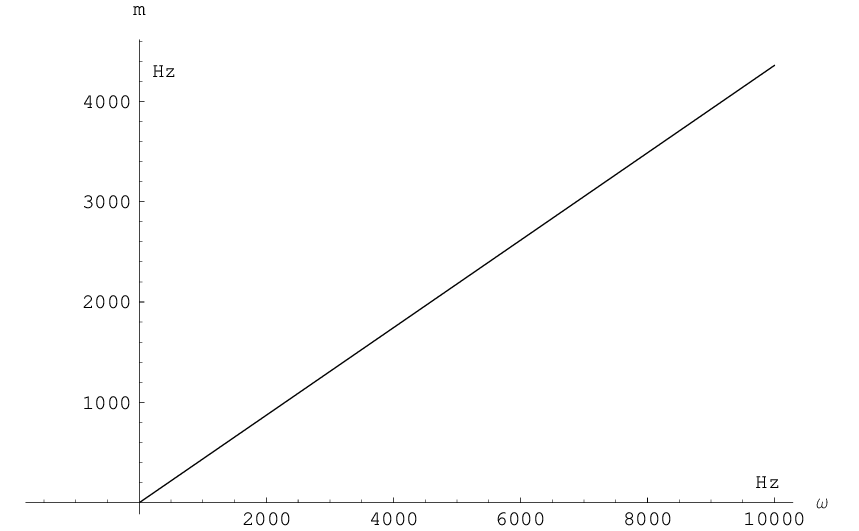}

\caption{the mass-frequency relation for a massive GW propagating with a speed
of $0.9c$ : for the mass it is $1Hz=10^{-15}eV$}
\end{figure}

Now, the analisys can remain in the Lorenz gauge with trasformations
of the type $\square\epsilon_{\nu}=0$; this gauge gives a condition
of transversality for the ordinary part of the field: $k^{\mu}A_{\mu\nu}=0$,
but does not give the transversality for the total field $h_{\mu\nu}$.
From eq. (\ref{eq: ridefiniz}) it is

\begin{equation}
h_{\mu\nu}=\bar{h}_{\mu\nu}-\frac{\bar{h}}{2}\eta_{\mu\nu}+\eta_{\mu\nu}h_{m}.\label{eq: ridefiniz 2}\end{equation}

At this point, if being in the massless case \cite{key-9,key-12,key-19},
it could been put

\begin{equation}
\begin{array}{c}
\square\epsilon^{\mu}=0\\
\\\partial_{\mu}\epsilon^{\mu}=-\frac{\bar{h}}{2}+h_{m},\end{array}\label{eq: gauge2}\end{equation}

which gives the total transversality of the field. But in the massive
case this is impossible. In fact, applying the Dalembertian operator
to the second of eqs. (\ref{eq: gauge2}) and using the field equations
(\ref{eq: onda T}) and (\ref{eq: onda S}) it results

\begin{equation}
\square\epsilon^{\mu}=m^{2}h_{m},\label{eq: contrasto}\end{equation}

which is in contrast with the first of eqs. (\ref{eq: gauge2}). In
the same way it is possible to show that it does not exist any linear
relation between the tensorial field $\bar{h}_{\mu\nu}$ and the massive
field $h_{m}$. Thus a gauge in wich $h_{\mu\nu}$ is purely spatial
cannot be chosen (i.e. it cannot be put $h_{\mu0}=0,$ see eq. (\ref{eq: ridefiniz 2}))
. But the traceless condition to the field $\bar{h}_{\mu\nu}$ can
be put :

\begin{equation}
\begin{array}{c}
\square\epsilon^{\mu}=0\\
\\\partial_{\mu}\epsilon^{\mu}=-\frac{\bar{h}}{2}.\end{array}\label{eq: gauge traceless}\end{equation}

These equations imply

\begin{equation}
\partial^{\mu}\bar{h}_{\mu\nu}=0.\label{eq: vincolo}\end{equation}

To save the conditions $\partial_{\mu}\bar{h}^{\mu\nu}$ and $\bar{h}=0$
transformations like

\begin{equation}
\begin{array}{c}
\square\epsilon^{\mu}=0\\
\\\partial_{\mu}\epsilon^{\mu}=0\end{array}\label{eq: gauge 3}\end{equation}

can be used and, taking $\overrightarrow{p}$ in the $z$ direction,
a gauge in which only $A_{11}$, $A_{22}$, and $A_{12}=A_{21}$ are
different to zero can be chosen. The condition $\bar{h}=0$ gives
$A_{11}=-A_{22}$. Now, putting these equations in eq. (\ref{eq: ridefiniz 2}),
it results

\begin{equation}
h_{\mu\nu}(t,z)=A^{+}(t-z)e_{\mu\nu}^{(+)}+A^{\times}(t-z)e_{\mu\nu}^{(\times)}+h_{m}(t-v_{G}z)\eta_{\mu\nu}.\label{eq: perturbazione totale}\end{equation}

The term $A^{+}(t-z)e_{\mu\nu}^{(+)}+A^{\times}(t-z)e_{\mu\nu}^{(\times)}$
describes the two standard polarizations of gravitational waves which
arise from General Relativity, while the term $h_{m}(t-v_{G}z)\eta_{\mu\nu}$
is the massive field arising from the high order theory. In other
words, the function $R^{\varepsilon}$ of the Ricci scalar generates
a third massive polarization for gravitational waves which is not
present in standard General Relativity.

\section{The primordial production of the third polarization}

Now, let us consider the primordial physical process, which gave rise
to a characteristic spectrum $\Omega_{gw}$ for the relic GWs. Such
physical process has been analyzed in different ways: respectively
in refs. \cite{key-25,key-26} and \cite{key-27,key-28} but only
for the components of eq. (\ref{eq: perturbazione totale}) which
arises from General Relativity, while in \cite{key-29} the process
has been extended to scalar-tensor gravity. Actually the process can
be furtherly improved showing the primordial production of the third
polarization of eq. (\ref{eq: perturbazione totale}).

Before starting with the analysis, it has to be emphasized that, considering
a stochastic background of GWs, it can be characterized by a dimensionless
spectrum \cite{key-25,key-26,key-27,key-28,key-29}\begin{equation}
\Omega_{gw}(f)\equiv\frac{1}{\rho_{c}}\frac{d\rho_{gw}}{d\ln f},\label{eq: spettro}\end{equation}

where \begin{equation}
\rho_{c}\equiv\frac{3H_{0}^{2}}{8G}\label{eq: densita' critica}\end{equation}

is the (actual) critical density energy, $\rho_{c}$ of the Universe,
$H_{0}$ the actual value of the Hubble expansion rate and $d\rho_{gw}$
the energy density of relic GWs in the frequency range $f$ to $f+df$.

The existence of a relic stochastic background of GWs is a consequence
of generals assumptions. Essentially it derives from a mixing between
basic principles of classical theories of gravity and of quantum field
theory. The strong variations of the gravitational field in the early
universe amplifie the zero-point quantum oscillations and produce
relic GWs. It is well known that the detection of relic GWs is the
only way to learn about the evolution of the very early universe,
up to the bounds of the Planck epoch and the initial singularity \cite{key-21,key-25,key-26,key-27,key-28,key-29}.
It is very important to stress the unavoidable and fundamental character
of this mechanism. The model derives from the inflationary scenario
for the early universe \cite{key-36,key-37}, which is tuned in a
good way with the WMAP data on the Cosmic Background Radiation (CBR)
(in particular exponential inflation and spectral index $\approx1$
\cite{key-38,key-39}). Inflationary models of the early Universe
were analysed in the early and middles 1980's (see \cite{key-36}
for a review ), starting from an idea of A. Guth \cite{key-37}. These
are cosmological models in which the Universe undergoes a brief phase
of a very rapid expansion in early times. In this context the expansion
could be power-law or exponential in time. Inflationary models provide
solutions to the horizon and flatness problems and contain a mechanism
which creates perturbations in all fields. Important for our goals
is that this mechanism also provides a distinctive spectrum of relic
GWs. The GWs perturbations arise from the uncertainty principle and
the spectrum of relic GWs is generated from the adiabatically-amplified
zero-point fluctuations \cite{key-21,key-25,key-26,key-27,key-28,key-29}. 

Now, the calculation for a simple inflationary model will be shown
for the third polarization of eq. (\ref{eq: perturbazione totale}),
following the works of Allen \cite{key-25,key-26} that performed
the calculation in the case of standard General Relativity and Corda,
Capozziello and De Laurentis \cite{key-29,key-30} that extended the
process to scalar GWs. Even here we have to recall that, in a recent
paper \cite{key-40}, such a process has been applied to the theory
arising from the action (\ref{eq: high order 1}). But, in \cite{key-40}
a different point of view has been considered. In that case, using
a conform analysis, the authors discussed such a process in respect
to the two standard polarizations which arises from standard General
Relativity. In the following the analysis is focused to the third
massive polarization. Thus, in a certain sense, one can say that the
present analysis is an integration of the analysis in \cite{key-40}.

It will be assumed that the universe is described by a simple cosmology
in two stages, an inflationary De Sitter phase and a radiation dominated
phase \cite{key-21,key-25,key-26,key-27,key-28,key-29}. The line
element of the spacetime is given by

\begin{equation}
ds^{2}=a^{2}(\eta)[-d\eta^{2}+d\overrightarrow{x}^{2}+h_{\mu\nu}(\eta,\overrightarrow{x})dx^{\mu}dx^{\nu}].\label{eq: metrica}\end{equation}

In this line element, because we are considering only the third polarization,
the metric perturbation (\ref{eq: perturbazione totale}) reduces
to 

\begin{equation}
h_{\mu\nu}=h_{m}I_{\mu\nu},\label{eq: perturbazione scalare}\end{equation}

where \begin{equation}
I_{\mu\nu}\equiv\begin{array}{cccc}
1 & 0 & 0 & 0\\
0 & 1 & 0 & 0\\
0 & 0 & 1 & 0\\
0 & 0 & 0 & 1.\end{array}\label{eq: identica}\end{equation}

In the De Sitter phase ($\eta<\eta_{1}$) the equation of state is
$P=-\rho=const$, the scale factor is $a(\eta)=\eta_{1}^{2}\eta_{0}^{-1}(2\eta_{1}-\eta)^{-1}$
and the Hubble constant is given by $H(\eta)=H_{ds}=c\eta_{0}/\eta_{1}^{2}$.

In the radiation dominated phase $(\eta>\eta_{1})$ the equation of
state is $P=\rho/3$, the scale factor is $a(\eta)=\eta/\eta_{0}$
and the Hubble constant is given by $H(\eta)=c\eta_{0}/\eta^{2}$~\cite{key-21,key-25,key-26,key-29,key-30}.

Expressing the scale factor in terms of comoving time defined by

\begin{equation}
cdt=a(t)d\eta\label{eq: tempo conforme}\end{equation}

one gets

\begin{equation}
a(t)\propto\exp(H_{ds}t)\label{eq: inflazione}\end{equation}

during the De Sitter phase and

\begin{equation}
a(t)\propto\sqrt{t}\label{eq: dominio radiazione}\end{equation}

during the radiation dominated phase. In order to obtain a solution
for the horizon and flatness problems it has to be \cite{key-36,key-37}

\begin{center}$\frac{a(\eta_{0})}{a(\eta_{1})}>10^{27}$\end{center}

The third polarization generates weak perturbations $h_{\mu\nu}(\eta,\overrightarrow{x})$
of the metric (\ref{eq: perturbazione scalare}) that can be written
in the form 

\begin{equation}
h_{\mu\nu}=I_{\mu\nu}(\hat{k})X(\eta)\exp(\overrightarrow{k}\cdot\overrightarrow{x}),\label{eq: relic gravity-waves}\end{equation}

in terms of the conformal time $\eta$ where $\overrightarrow{k}$
is a constant wavevector and

\begin{equation}
h_{m}(\eta,\overrightarrow{k},\overrightarrow{x})=X(\eta)\exp(\overrightarrow{k}\cdot\overrightarrow{x}).\label{eq: phi}\end{equation}
By putting $Y(\eta)=a(\eta)X(\eta)$  and with the standard linearized
calculation in which the connections (i.e. the Cristoffel coefficents),
the Riemann tensor, the Ricci tensor and the Ricci scalar curvature
are found, from Friedman linearized equations it is obtained that
the function $Y(\eta)$ satisfies the equation

\begin{equation}
Y''+(|\overrightarrow{k}|^{2}-\frac{a''}{a})Y=0\label{eq: Klein-Gordon}\end{equation}

where $'$ denotes derivative with respect to the conformal time.
Cleary, this is the equation for a parametrically disturbed oscillator.

The solutions of eq. (\ref{eq: Klein-Gordon}) give us the solutions
for the function $X(\eta)$, that can be expressed in terms of elementary
functions simple cases of half integer Bessel or Hankel functions
\cite{key-21,key-25,key-26,key-29,key-30} in both the inflationary
and radiation dominated eras:

For $\eta<\eta_{1}$ \begin{equation}
X(\eta)=\frac{a(\eta_{1})}{a(\eta)}[1+H_{ds}\omega^{-1}]\exp-ik(\eta-\eta_{1}),\label{eq: ampiezza inflaz.}\end{equation}

for $\eta>\eta_{1}$ \begin{equation}
X(\eta)=\frac{a(\eta_{1})}{a(\eta)}[\alpha\exp-ik(\eta-\eta_{1})+\beta\exp ik(\eta-\eta_{1}),\label{eq: ampiezza rad.}\end{equation}

where $\omega=ck/a$ is the angular frequency of the wave (that is
function of the time because of the constance of $k=|\overrightarrow{k}|$),
$\alpha$ and $\beta$ are time-indipendent constants which can be
obtained demanding that both $X$ and $dX/d\eta$ are continuous at
the boundary $\eta=\eta_{1}$ between the inflationary and the radiation
dominated eras of the cosmologic expansion. With this constrain it
is

\begin{equation}
\alpha=1+i\frac{\sqrt{H_{ds}H_{0}}}{\omega}-\frac{H_{ds}H_{0}}{2\omega^{2}}\label{eq: alfa}\end{equation}

\begin{equation}
\beta=\frac{H_{ds}H_{0}}{2\omega^{2}}\label{eq: beta}\end{equation}

In eqs. (\ref{eq: alfa}), (\ref{eq: beta}) $\omega=ck/a(\eta_{0})$
is the angular frequency that would be observed today. Calculations
like this are referred in the literature as Bogoliubov coefficient
methods \cite{key-21,key-25,key-26,key-29,key-30}. 

As inflation dampes out any classical or macroscopic perturbations,
the minimum allowed level of fluctations is that requiered by the
uncertainty principle. The solution (\ref{eq: ampiezza inflaz.})
corresponds precisely to this De Sitter vacuum state \cite{key-21,key-25,key-26,key-29,key-30}.
Then, if the period of inflation was long enough, the observable properties
of the Universe today should be the same properties of a Universe
started in the De Sitter vacuum state.

In the radiation dominated phase the coefficients of $\alpha$ are
the eigenmodes which describe particles while the coefficients of
$\beta$ are the eigenmodes which describe antiparticles. Thus, the
number of created particles of angular frequency $\omega$ in this
phase is 

\begin{equation}
N_{\omega}=|\beta_{\omega}|^{2}=(\frac{H_{ds}H_{0}}{2\omega^{2}})^{2}.\label{eq: numero quanti}\end{equation}

Now, one can write an expression for the energy spectrum of the relic
gravitational waves background in the frequency interval $(\omega,\omega+d\omega)$
as

\begin{equation}
d\rho_{gw}=2\hbar\omega(\frac{\omega^{2}d\omega}{2\pi^{2}c^{3}})N_{\omega}=\frac{\hbar H_{ds}^{2}H_{0}^{2}}{4\pi^{2}c^{3}}\frac{d\omega}{\omega}=\frac{\hbar H_{ds}^{2}H_{0}^{2}}{4\pi^{2}c^{3}}\frac{df}{f}.\label{eq: de energia}\end{equation}

Eq. (\ref{eq: de energia}) can be rewritten in terms of the present
day and the De Sitter energy-density of the Universe. The Hubble expansion
rates is

\begin{center}$H_{0}^{2}=\frac{8\pi G\rho_{c}}{3c^{2}}$, $H_{ds}^{2}=\frac{8\pi G\rho_{ds}}{3c^{2}}$.\end{center}

Then, defining the Planck density \begin{equation}
\rho_{Planck}\equiv\frac{c^{7}}{\hbar G^{2}}\label{eq: Shoooortyyyyyy}\end{equation}
 the spectrum is 

\begin{equation}
\Omega_{gw}(f)=\frac{1}{\rho_{c}}\frac{d\rho_{sgw}}{d\ln f}=\frac{f}{\rho_{c}}\frac{d\rho_{gw}}{df}=\frac{16}{9}\frac{\rho_{ds}}{\rho_{Planck}}.\label{eq: spettro gravitoni}\end{equation}

Some comments are needed. It has to be emphasized that the computation
works for a very simplified model that does not include the matter
dominated era. Including this era, the redshift has to be considered.
An enlighting computation parallel to the one in \cite{key-26} gives

\begin{equation}
\Omega_{gw}(f)=\frac{16}{9}\frac{\rho_{ds}}{\rho_{Planck}}(1+z_{eq})^{-1},\label{eq: spettro gravitoni redshiftato}\end{equation}

for the waves which at the time in which the Universe was becoming
matter dominated had a frequency higher than $H_{eq}$, the Hubble
constant at that time. This corresponds to frequencies $f>(1+z_{eq})^{1/2}H_{0}$,
where $z_{eq}$ is the redshift of the Universe when the matter and
radiation energy density were equal. The redshift correction in equation
(\ref{eq: spettro gravitoni redshiftato}) is needed because the Hubble
parameter, which is governed by Friedman equations, should be different
from the observed one $H_{0}$ for a Universe without matter dominated
era.

At lower frequencies the spectrum is \cite{key-21,key-25,key-26,key-29,key-30}

\begin{equation}
\Omega_{gw}(f)\propto f^{-2}.\label{eq: spettro basse frequenze}\end{equation}

Moreover, let us note that the results (\ref{eq: spettro gravitoni})
and (\ref{eq: spettro gravitoni redshiftato}), which are not frequency
dependent, cannot be applied to all the frequencies. For waves with
frequencies less than $H_{0}$ today, the energy density cannot be
defined, because the wavelenght becomes longer than the Hubble radius.
In the same way, at high frequencies there is a maximum frequency
above which the spectrum drops to zero rapidly. In the above computation
it has been implicitly assumed that the phase transition from the
inflationary to the radiation dominated epoch is istantaneous. In
the real Universe this phase transition occurs over some finite time
$\Delta\tau$, and above a frequency

\begin{equation}
f_{max}=\frac{a(t_{1})}{a(t_{0})}\frac{1}{\Delta\tau},\label{eq: freq. max}\end{equation}

which is the redshifted rate of the transition, $\Omega_{gw}$ drops
rapidly. These two cutoffs, at low and high frequencies, to the spectrum
force the total energy density of the relic gravitational waves to
be finite. For GUT energy-scale inflation it is \cite{key-21,key-25,key-26,key-29,key-30}. 

\begin{equation}
\frac{\rho_{ds}}{\rho_{Planck}}\approx10^{-12}.\label{eq: rapporto densita' primordiali}\end{equation}

\section{Tuning with WMAP data}

It is well known that WMAP observations put strongly severe restrictions
on the spectrum of relic gravitational waves. In fig. 2 the spectrum
$\Omega_{gw}$is mapped following \cite{key-20}: the amplitude is
chosen (determined by the ratio $\frac{\rho_{ds}}{\rho_{Planck}}$)
to be \textit{as large as possible, consistent with the WMAP constraints}
o\textit{n tensor perturbations.} Nevertheless, because the spectrum
falls off $\propto f^{-2}$ at low frequencies, this means that today,
at LIGO-Virgo and LISA frequencies (indicate by the lines in fig.
2) \cite{key-20}, it is

\begin{center}\begin{equation}
\Omega_{gw}(f)h_{100}^{2}<9*10^{-13}.\label{eq: limite spettro WMAP}\end{equation}
\end{center}

It is interesting to calculate the correspondent strain at $\approx100Hz$,
where interferometers like Virgo and LIGO have a maximum in sensitivity.
The well known equation for the characteristic amplitude, adapted
for the third component of GWs can be used \cite{key-20}:

\begin{equation}
h_{mc}(f)\simeq1.26*10^{-18}(\frac{1Hz}{f})\sqrt{h_{100}^{2}\Omega_{gw}(f)},\label{eq: legame ampiezza-spettro}\end{equation}
obtaning \cite{key-20}

\begin{equation}
h_{mc}(100Hz)<1.7*10^{-26}.\label{eq: limite per lo strain}\end{equation}

Then, as we expect a sensitivity of the order of $10^{-22}$ for our
interferometers at $\approx100Hz$, we need to gain four order of
magnitude. Let us analyze smaller frequencies too. The sensitivity
of the Virgo interferometer is of the order of $10^{-21}$ at $\approx10Hz$
and in that case it is \cite{key-20}

\begin{equation}
h_{mc}(10Hz)<1.7*10^{-25}.\label{eq: limite per lo strain2}\end{equation}

The sensitivity of the LISA interferometer will be of the order of
$10^{-22}$ at $10^{-3}\approx Hz$ and in that case it is \cite{key-20}

\begin{equation}
h_{mc}(100Hz)<1.7*10^{-21}.\label{eq: limite per lo strain3}\end{equation}

Then, a stochastic background of relic gravitational waves could be
in principle detected by the LISA interferometer.

We emphasize that the assumption that all the tensorial perturbation
in the Universe are due to a stochastic background of GWs is quit
strong, but our results (\ref{eq: limite spettro WMAP}), (\ref{eq: limite per lo strain}),
(\ref{eq: limite per lo strain2}) and (\ref{eq: limite per lo strain3})
can be considered like upper bounds.

\begin{figure}[H]
\includegraphics[%
  scale=0.9]{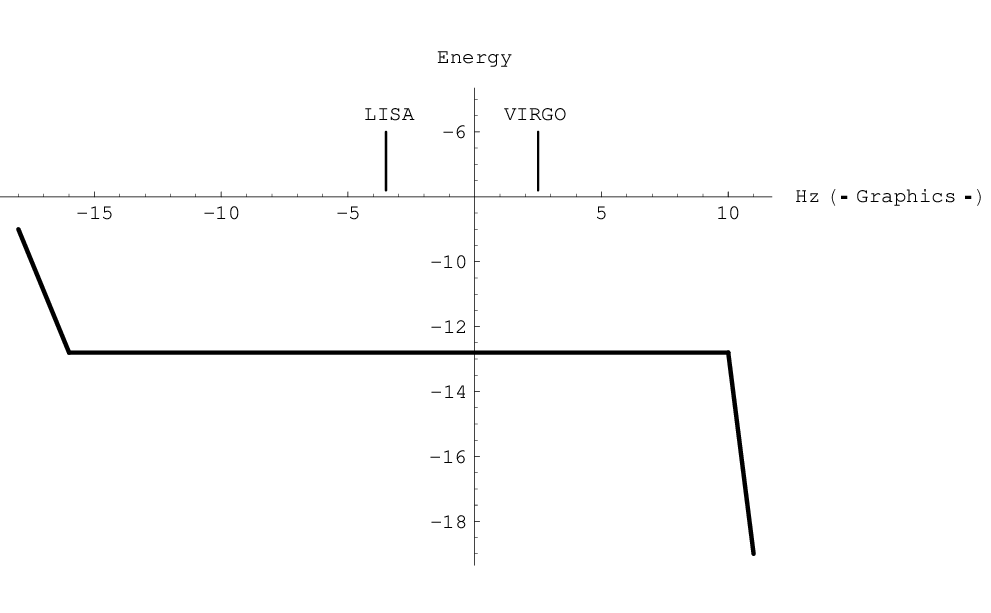}

\caption{adapted from C. Corda - Mod. Phys. Lett. A No. 22, 16, 1167-1173
(2007). }

The spectrum of relic sGWs in inflationary models is flat over a wide
range of frequencies. The horizontal axis is $\log_{10}$ of frequency,
in Hz. The vertical axis is $\log_{10}\Omega_{gsw}$. The inflationary
spectrum rises quickly at low frequencies (wave which rentered in
the Hubble sphere after the Universe became matter dominated) and
falls off above the (appropriately redshifted) frequency scale $f_{max}$
associated with the fastest characteristic time of the phase transition
at the end of inflation. The amplitude of the flat region depends
only on the energy density during the inflationary stage; we have
chosen the largest amplitude consistent with the WMAP constrains on
scalar perturbations. This means that at LIGO and LISA frequencies,
$\Omega_{gw}(f)h_{100}^{2}<9*10^{-13}$
\end{figure}

\section{Conclusions}

It has been shown that a weak modification of general relativity produces
a third massive polarization of gravitational waves and the primordial
production of this polarization has been analysed adapting the adiabatically-amplified
zero-point fluctuations process at this case. 

The presence of the mass could also have important applications in
cosmology because the fact that gravitational waves can have mass
could give a contribution to the dark matter of the Universe.

At the end of the paper, an upper bound for these relic gravitational
waves, which arises from the WMAP constrains, has also been released
.

\section*{Acknowledgements}

I would like to thank Salvatore Capozziello and Maria Felicia De Laurentis
for helpful advices during my work. The European Gravitational Observatory
(EGO) consortium has also to be thanked for the using of computing
facilities.

\end{document}